\documentclass[aps,prl,twocolumn]{revtex4-1}
\usepackage{graphics,color,graphicx,amsmath}
\usepackage{xr,natbib}
\usepackage{amsbsy}

\graphicspath{{images}}

\begin{document}

\newcommand{\mr}[1]{\mathrm{#1}}
\newcommand{\mb}[1]{\mathbf{#1}}
\newcommand{\br}[1]{\left<#1\right>}
\newcommand{\bl}[1]{\left|#1\right|}
\newcommand{\mc}[1]{\mathcal{#1}}
\newcommand{\tb}[1]{\textcolor{blue}{#1}}
\newcommand{\tr}[1]{\textcolor{red}{#1}}
\newcommand{\tg}[1]{\textcolor{green}{#1}}
\newcommand{\hi}[0]{h_{\rm i}}
\newcommand{\Jij}[0]{J_{\rm ij}}
\newcommand{\si}[0]{s_{\rm i}}
\newcommand{\sj}[0]{s_{\rm j}}
\newcommand{\sk}[0]{s_{\rm k}}
\newcommand{\qi}[0]{q_{\rm i}}
\newcommand{\bs}[1]{\boldsymbol{#1}}
\newcommand{\mN}[0]{\mathcal{N}}

\title{Closely estimating the entropy of sparse graph models}
\author{Edward D.~Lee}
\affiliation{Complexity Science Hub Vienna, Josefst\ae dter Strasse 39, Vienna, Austria}

\begin{abstract}
We introduce an algorithm for estimating the entropy of pairwise, probabilistic graph models by leveraging bridges between social communities and an accurate entropy estimator on sparse samples. We propose using a measure of investment from the sociological literature, Burt's structural constraint, as a heuristic for identifying bridges that partition a graph into conditionally independent components. We combine this heuristic with the Nemenman-Shafee-Bialek entropy estimator to obtain a faster and more accurate estimator. We demonstrate it on the pairwise maximum entropy, or Ising, models of judicial voting, to improve na\"{i}ve entropy estimates. We use our algorithm to estimate the partition function closely, which we then apply to the problem of model selection, where estimating the likelihood is difficult. This serves as an improvement over existing methods that rely on point correlation functions to test fit can be extended to other graph models with a straightforward modification of the open-source implementation.
\end{abstract}

\maketitle

Estimating the entropy of a probabilistic model is a common and essential yet difficult task in information theoretic analysis of collective behavior. For example, widely used maximum entropy models of neural firing \cite{schneidmanWeakPairwise2006, tkacikIsingModels2006,bartonIsingModels2013, chenSearchingCollective2018, leeDiscoveringSparse2022}, political voting \cite{leeStatisticalMechanics2015}, social groups \cite{hallStatisticalMechanics2019, leeSensitivityCollective2020}, or collective motion \cite{shemeshHighorderSocial2013,bialekStatisticalMechanics2012} require an estimate of the entropy of the model for assessing the fit with the multi-information. Furthermore, the relative decrease in the entropy from the independent model provides a general metric for collectivity, or a distance from independent statistics, that gives a sense of the constraints imposed by model assumptions. The entropy is intimately connected to the free energy, which describes the balance between energy constraints and disorder and thus can be used to assess the goodness of fit. Finally, the free energy is the logarithm of the partition function, the derivatives of which reveal physical properties such as the magnetization, susceptibility, and the distance to a critical point \cite{reifFundamentalsPhysics1965,danielsControlFinite2017}.

The general problem of entropy estimation even when given a model is hard. Statistical physics approaches including moment expansions like the Bethe free energy approximation \cite{yedidiaBetheFree2001,mezardInformationPhysics2009} and cluster expansions \cite{coccoAdaptiveCluster2011} provide one approach along with thermodynamic integration \cite{frenkelUnderstandingMolecular2002}. An alternative approach is to infer the entropy from a sample of the distribution, which can be relatively quick to generate using Monte Carlo Markov chain methods. Most interesting systems, however, are of sufficient size that the state space dwarfs the number of possible (or reasonable) samples from simulation, a problem that is especially intractable in the space of discrete outcomes. Unfortunately, we know well that estimates of the entropy in the undersampled limit are biased. Even worse, starting with a prior in the space of probability distributions in order to infer a good estimate of entropy can backfire, introducing other biases that cannot be overcome in the low data limit. A remarkably accurate estimator, called NSB after the authors Nemenman-Shafee-Bialek \cite{nemenmanEntropyInference2002}, instead attempts to be as unbiased in the estimate by using a prior that is flat in the space of entropies and as a result can perform well in the highly undersampled limit. Like any other estimator, however, it will still face problems in larger systems because of the prior inevitably dominates and (in practical terms) the computation is slower and numerically challenging. 

Here, we are interested in the problem of estimating the entropy when already given a model and leveraging entropy estimators from small samples in the literature like NSB such as when characterizing the statistics of natural phenomena \cite{strongEntropyInformation1998}. In short, we address a technical problem in the numerical estimation of the entropy of a probabilistic model leveraging the fact that many problems show sparse structure. In the context of physical models, this is the observation that components can be split into communities that are sparsely connected to one another. This the case for many political interaction graphs such as for judicial courts or for legislatures, where the vast majority of occupants over time have never met or interacted with one another \cite{leePartisanIntuition2018}. This is also the case for some social graphs which tend to be split into cliques that are connected only indirectly to one another in small-world networks \cite{burtStructuralHoles1992,newmanWhySocial2003,wattsCollectiveDynamics1998}, but it is not the case for small animal groups that display lattice-like, topologically local interactions such as in bird flocks and local interaction models for herding behavior \cite{balleriniInteractionRuling2008}. 
As we describe below, we develop a heuristic that leverages existing estimators and package it into a Python library TreeEnt that can estimate faster and more accurately the entropy of sparsely connected graph models in a discrete state space.

\section{Entropy \& its estimation in a nutshell}
In the mid 20th century, Claude Shannon sought a way to characterize the statistical structure of signals sent through AT\&T's telecommunication networks \cite{gleickInformationHistory2011}. By treating a message as a sequence of discrete characters $s$ each occurring with a probability $p(s)$, he established the ``information entropy'' measuring the surprise that each new character in the received sequence would entail. He furthermore showed that this was a unique measure (barring a choice of units) adhering to three axioms \cite{shannonMathematicalTheory1948}: continuity with respect to $p$, maximization at maximal uncertainty, and consistency under a hierarchical decomposition of the symbols into sets. Formally, the information entropy $H$ of a probability distribution $p(s)$ for a configuration $s$ from a discrete-state space $\mathcal S$ is \cite{coverElementsInformation2006} 
\begin{align}
	H[p] &\equiv -\sum_{s\in\mathcal S} p(s) \log p(s)\label{eq:H}.
\end{align}
The entropy is maximized when the distribution over all configurations $s$ is uniform (and thus completely unpredictable) and zero when only a single configuration occurs with probability one (completely predictable). Thus, the entropy presents a unique measure of the amount of structure in the distribution and places limits on its {\it statistical} predictability.

Estimation of the entropy given a random sample from $p(s)$ is difficult because the state space is computationally expensive, if not impossible, to sample comprehensively even for systems of moderate size. If we, for example, try estimating the entropy in the most na\"ive way; that is, to rely on the number of times $k_s$ that we see state $s$ from a random, finite sample of size $K$, we would posit that $\hat p(s) = k_s/K$. Then, the deviation from the true value can be represented as an error term, $p(s) = \hat p(s) + \epsilon(s)$. Expanding the entropy in terms of the errors, we find
\begin{align}
	\hat H[p] &= H[p] - \frac{A}{K} - \mathcal{O}\left[\frac{1}{K^2}\right],
\end{align}
where we have grouped into the last term all terms of order $(1/K)^2$ and higher and $A$ is a positive constant. Thus, it is the case that the na\"ive estimator will return a biased estimate that underpredicts the true entropy of the distribution.

\begin{figure}\centering
	\includegraphics[width=\linewidth]{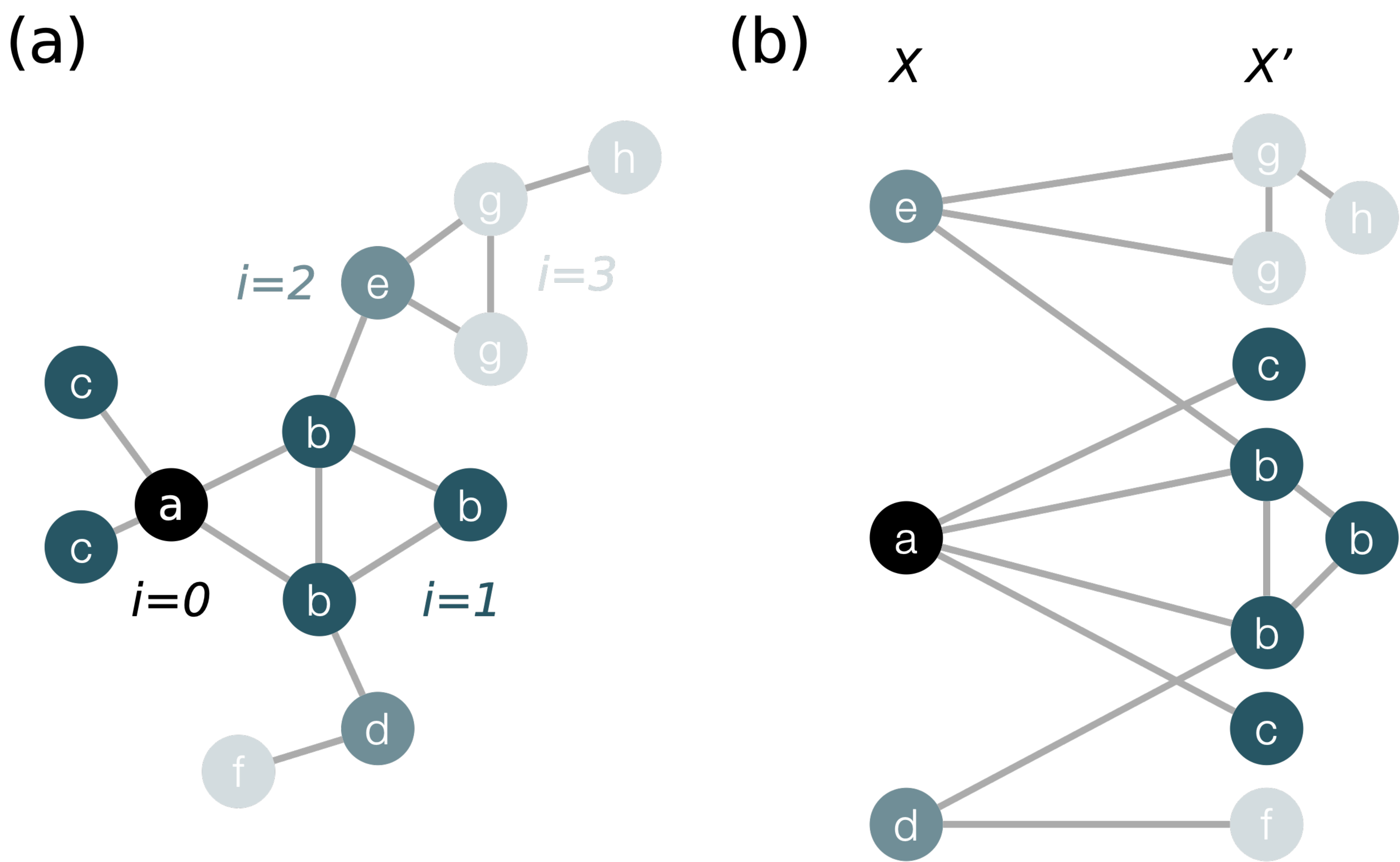}
	\caption{Example of graph partition. (a) Graph before partition. Edges denote interactions between nodes. We index with increasing index i each generation of the tree away from the root at $\rm i=0$. Generations also differentiated by shading. When each branch is indexed by the alphabet, we specify the set of nodes by the generation and branch index, e.g.~$\si^a$. (b) Bipartite graph from our partition algorithm. Downstream nodes (e.g.~b relative to a) are conditionally independent once fixing the values of upstream nodes. This permits us to reduce the sample state space exponentially because the largest subgraph determines the largest state space to sample. Graph partitioned by setting maximum cluster size to $K=4$.}\label{gr:graph}
\end{figure}

One way to ameliorate this problem may be a Bayesian approach in which we define entropy estimation as an inference task over a prior distribution of the probability distributions from which the sample originates. To be pedantic, a classic way to do this is to write a Dirichlet family of models where we specify the probabilities $\qi$ with which each unique state occurs out of the possible set of $S$ states (e.g.~for a binary spin system of size $N$, we would have $S=2^N$). Accounting for normalization of the set of probabilities $\{\qi\}$, we have
\begin{align}
	P_\beta(\{\qi\}) &= \frac{1}{Z(\beta)} \delta \left( 1-\sum_{\rm i=1}^S \qi \right) \prod_{\rm i=1}^S\qi^{\beta-1}.\label{eq:dirichlet prior}
\end{align}
This is known as the Dirichlet family of priors where $\beta$, which determines how likely we consider larger probabilities {\it a priori}; Laplace's counting rule corresponds to $\beta=1$ and the maximum likelihood estimator $\beta=0$. The factor $Z$ ensures that this is a normalized probability distribution. Given the prior in Eq~\ref{eq:dirichlet prior}, the distribution over the estimated entropy $H$ will be given by Bayes' theorem,
\begin{align}
	P_\beta(H) &= \frac{P_\beta(H|\{\qi\})P_\beta(\{\qi\})}{P_\beta(\{\qi\}|H)}.
\intertext{Then, the averaged estimate of the entropy given $\beta$ is}
	\xi(\beta) &\equiv \int_0^{\infty} P_\beta(H) H\,dH.
\end{align}
Crucially, the choice of $\beta$ plays a crucial role in determining the final estimate of then entropy because the prior is peaked and thus essentially determines the entropy in the low data sample size limit \cite{nemenmanEntropyInference2002}. 

Instead, we might consider a meta-prior that consider a mixture of all possible priors within this family, or the range of possible $\beta$ in a way that flattens our starting assumptions about the entropy. This means that we should move $\beta$ over some weighted range $P(\beta)$ in such a way that corresponds to moving $\xi(\beta)$ between 0 and $\log H$. Then, a candidate mixture prior proposed in reference \citenum{nemenmanEntropyInference2002} is
\begin{align}
	P(\{\qi\}; \beta) &= \frac{1}{Z(\beta)} \delta\left( 1-\sum_{\rm i=1}^S \qi \right) \prod_{\rm i=1}^S \qi^{\beta-1} \frac{d\xi(\beta)}{d\beta} P(\beta).\label{eq:nsb prior}
\end{align}
To perform this integral, one must also know how the estimated entropy varies as a function of $\xi$ in order to flatten the prior in the estimated range, the relationship between which is given in reference \citenum{nemenmanEntropyInference2002}. As it turns out, Eq~\ref{eq:nsb prior} performs surprisingly well for entropy estimates on tiny samples. Regardless, it cannot overcome the fundamental problem that entropy estimates are dominated by the prior in any undersampled system.

One way out of this pickle is to leverage the structure of the model's probability distribution to simplify the entropy estimation problem by effectively reducing the state space over which one's estimator needs to work. Conveniently, Shannon's last axiom tells us that if we can group the states into independent subsets, then the information entropy is the sum of both the uncertainty of the labels from the clusters as well as the contribution from within each subset given the labels. When a clever decomposition is possible, it would allow us to consider subsets of the system independently of one another. This, for example, is possible if some components of $s$ with index i, the set $\{\si\}$ of which we denote $\si$ for simplicity, are conditionally independent of components indexed by j once holding fixed components k. In other words, this is the assertion that we can factorize the probability distribution
\begin{align}
	p(s) &= p(\si|\sk)p(\sj|\sk)p(\sk).
\end{align}
If this is the case, then the entropy of this set decomposes in the summation of the information entropies
\begin{align}
	H[p] &= H[p(\sk)] - \sum_{\sk} p(\sk) \left(H[p(\si|\sk)] + \right.\notag\\
		&\qquad\qquad\qquad\qquad\qquad\qquad    \left. H[p(\sj|\sk)]\right).
\end{align}
We can think of each unique configuration of $\sk$ as a ``label,'' and once this is given we must compute the uncertainty of the sets $\si$ and $\sj$ with their respective weights given by the frequency of the $\sk$ on which they have been conditioned.

When we generalize this basic example to a tree, we denote a set again as $\si$ but now denote moving up generation in the tree as the index $\rm i-1$ and moving down a generation as $\rm i+1$. Labeling each branch with a different letter of the alphabet as superscript,
\begin{align}
	p(s) &= p(s_0) \prod_{i=1}^N \prod_{a,a'} p\Big(s^{a'}_{\rm i}\Big|s_{\rm i-1}^{a}\Big),
\end{align}
where we again use the shorthand notation $\si$ to refer to the set of components over index i. Then, the root is the set of nodes $s_0$, the first product is over each successive layer in the tree indexed i down to the leaves in the $N$th generation, and the second product over the descendents $a'$ of each branch $a$ in the ($\rm i-1$)th layer, which is independent once having conditioned on all parent branches. As an example of such a factorization using this notation, we show a graph in Figure~\ref{gr:graph}, where we indicate the successive generation of the tree i generations away from the root that are conditionally independent of one another once conditioning on the parent branches $\rm i-1$. 

When the probability distribution is factorizable in this way, then it is possible to compute the entropy of the entire system as a sum of entropies from the outside in by computing the entropy of each leaf, which provides an additive contribution to the final entropy. We express this in the recursive form
\begin{widetext}
\begin{align}
	H[p] &= H[p(s_0)] + \sum_{s_0} \sum_{a} p(s_0) \Big( H[p(s_{1}^{a}|s_0)] + \sum_{s_1^a} \sum_{a'} p(s_1^{a}|s_0) \Big(H[p(s_{2}^{a'}|s_1^a)] + \cdots + \notag\\
		& \qquad\qquad\qquad\qquad\qquad\qquad \sum_{s_{i}^{a^{(n-1)}}} \sum_{a^{(n)}} p\left(s_i^{a^{(n)}}|s_{i-1}^{a^{(n-1)}}\right) \Big(H\left[p\left(s_{i+1}^{a^{(n+1)}}|s_i^{a^{(n)}}\right)\right] + \cdots \Big)\Big)\Big).\label{eq:recursive H}
\end{align}
\end{widetext}
In short, we can implement the calculation in Eq~\ref{eq:recursive H} from the leaves up to the root. We start with the conditional entropy of a leaf, prune the leaf, upon which the branch leading to the leaf in turn becomes a leaf. Continuing in this recursive way, we eventually reach the root of the graph. 

Such a hierarchical decomposition presents a way to ameliorate the problem of entropy estimation if we can group components into subsets that are substantially smaller than the full graph, which shrinks the state space exponentially \footnote{In a similar sense, one can also show that the complexity of the partition function for the Ising model on a square lattice does not go as $2^N$ but rather as $2^L$, where $L$ is the length of one side of the lattice, because one can condition on a line of spins that cuts the problem in half and do this recursively.}. If it is the case that the groups are substantially smaller than the full graph, estimating the entropy in principles becomes more manageable with an exponentially smaller Monte Carlo Markov chain (MCMC) sample. We present an algorithm that estimates the entropy of probabilistic graph models that factorizes the graph using a heuristic based on structural holes and fast MCMC sampling of the conditioned probability distributions.


\section{Algorithm}
We provide an outline of algorithm in Table~\ref{tab:algo} and provide more details below.

The first step is to compute a factorization of the graph that allows us to split the problem into more manageable entropy estimation problems. 
To do so, we rely on the notion of structural holes from the sociological literature \cite{burtStructuralHoles1992,burtStructuralHoles2004}. Structural holes are inspired by a problem of constrained action in a social network, where the assumption is that the amount that any individual $u$ is constrained by a neighbor $v$ depends directly on two factors: the relative investment that $u$ dedicates in its the relationship with $v$ and the simultaneous relative investment of another neighbor $w$ of $u$ into $v$. The intuition underlying the latter step is that when investments overlap, $u$ is maximally constrained in terms of leverage because $u$ is redundantly influencing the local neighborhood of $v$ (they are competing for influence in the same sphere), whereas with minimal overlap, $v$ acts less as a constraint and more as a bridge to other parts of the network to which $u$ does not have access. In other words, a neighbor $v$ with low structural constraint with respect to $u$ is surrounded by ``structural holes'' such as when it belongs to a different clique.

For our purposes, the intuition is that the nodes with low constraint are ones that are placed in between densely connected communities; they should present a small set that we can fix to render adjacent communities independent of one another. More specifically, these nodes are surrounded by holes and thus have small structural constraint $c_u$, defined as the sum over local constraints $l$ for node $u$,
\begin{align}
	c_u &:= \sum_{v \in \mathcal{N}(u)} l(u,v),\\
\intertext{which is the sum over all neighbors $v$ of $u$, or $\mN(u)$. The local constraint given uniform edge weights is defined as}
	l(u,v) &:= \left( \frac{1}{|\mN(u)|} + \sum_{w\in \mN(u)\cap\mN(v)} \frac{1}{|\mN(u)|}\frac{1}{|\mN(w)|} \right)^2, \label{eq:l}
\end{align}
where the number of neighbors of node $u$ is $|\mN(u)|$. The first term in the parentheses accounts for $u$'s investment in this neighbor $v$, which is uniform across all neighbors. The second term accounts for the joint investment in neighbor $v$ from both $u$ and a common neighbor $w$.  Note that Eq~\ref{eq:l} is a simplified form for our scenario, where ``investment,'' or the weights, between pairs of nodes does not generally need to be uniform \footnote{Burt proposes another measure for structural constraint, where instead the last term in Eq~\ref{eq:l} goes with the weight that $w$ puts on $v$ normalized by the maximum edge weight for $w$. Since the weights are uniform here, this would mean that the term becomes unity \cite{burtStructuralHoles1992}.}.

We start by removing nodes from the graph starting with the node of minimal constraint and recomputing the local constraints upon every removal step. Every removed node is placed into the set $X'$. This will tend to split apart connected components in the set of remaining nodes $X$ along the bridges that connected them. If we continue indefinitely with this procedure, the subgraphs living in the set of removed nodes $X'$ will eventually consist of most of the graph, and they may form large components that have not simplified the problem at all. Thus, we keep removing nodes with minimal constraint as long as the largest components in the set of removed nodes as well as the pruned graph are shrinking or until they have all fallen below a size smaller than the specified threshold $M$.

\begin{table}
\caption{Algorithm for entropy estimation. In the case the starting graph consists of multiple components, they are treated separately.}\label{tab:algo}
\begin{enumerate}
	\item Factorize graph to generate ``contracted'' tree.\\
	\subitem a. Place all nodes into $X$.
	\subitem b. Calculate the structual constraint for every node in $X$.
	\subitem c. Remove from set $X$ and place into $X'$ the node with minimal constraint in $X$.
	\subitem d. If the largest connected component in either $X$ or $X'$ is now larger than largest component previous to step c or if the largest connected component in $X$ is smaller than or equal to threshold $M$, then jump to step f.
	\subitem e. Return to step b.
	\subitem f. Create a coarse-grained representation of the graph, where a connected component in $X$ and $X'$ are connected to one another if there is at least one interaction between the two components.
	\item Estimate conditional entropies.\\
		\subitem a. Identify leaves in coarse-grained graph.
		\subitem b. For each leaf, generate a Monte Carlo sample of size $K$.
		\subitem c. For each unique state in the sample of size $K$, generate from the leaf a Monte Carlo sample of size $K'$.
		\subitem d. Estimate the entropy of the leaf and the conditioned branch using NSB.
		\subitem e. Prune the coarse-grained graph by removing all leaves.
		\subitem f. Return to first step.
	\item Sum entropies and calculate errors.
		\subitem a. Sum over the estimated entropies of each leaf and its branch.
\end{enumerate}
\end{table}

As a result of this procedure, we have two sets of nodes $X$ and $X'$. Within each set, we have disconnected components labeled $x$ and $x'$, respectively, that bridge nodes in the other but are not directly connected, or a bipartite structure as we show in Figure~\ref{gr:graph}b. Thus, the graph presents a structure in which fixing the components in $x$ renders components $x'$ independent of one another and vice versa.

As the final step, we calculate the entropy of the resulting factorized tree. We do so identifying a leaves, or components in either $X$ or $X'$ that are conditionally independent when holding a single component in the complementary set fixed. While we cannot guarantee that a leaf can be found at the end of our structural constraint removal procedure, it is the case that removing nodes of minimal constraint will tend to fragment the graph into a chain of conditionally independent pieces. Furthermore, there always exists a partition of the graph such that a leaf node can be designated (in the trivial case a single node can be put into $X'$). 

Finally, we then sample from the set of conditionally independent ``leaves'' and the unique set of nodes that lead to them, or their ``branch.'' First, we generate an MCMC sample of the distribution of the $a$'th branch $s_{\rm i-1}^{a}$ from a sample of the entire system. Then, we iterate through the unique states and sample for the $a'$'th leaf $\si^{a'}$ given each of the possible states of the branch. The size of the sample set also allows us to estimate the standard deviation of the sampled entropy. This gives us the terms in Eq~\ref{eq:recursive H}. Then, we prune the leaf from the graph and iterate this process recursively until we only have the root of the tree. At each step, we calculate the entropy using the NSB estimator. The total entropy of the tree is the summation of the calculated entropies.

\begin{figure}
	\includegraphics[width=\linewidth]{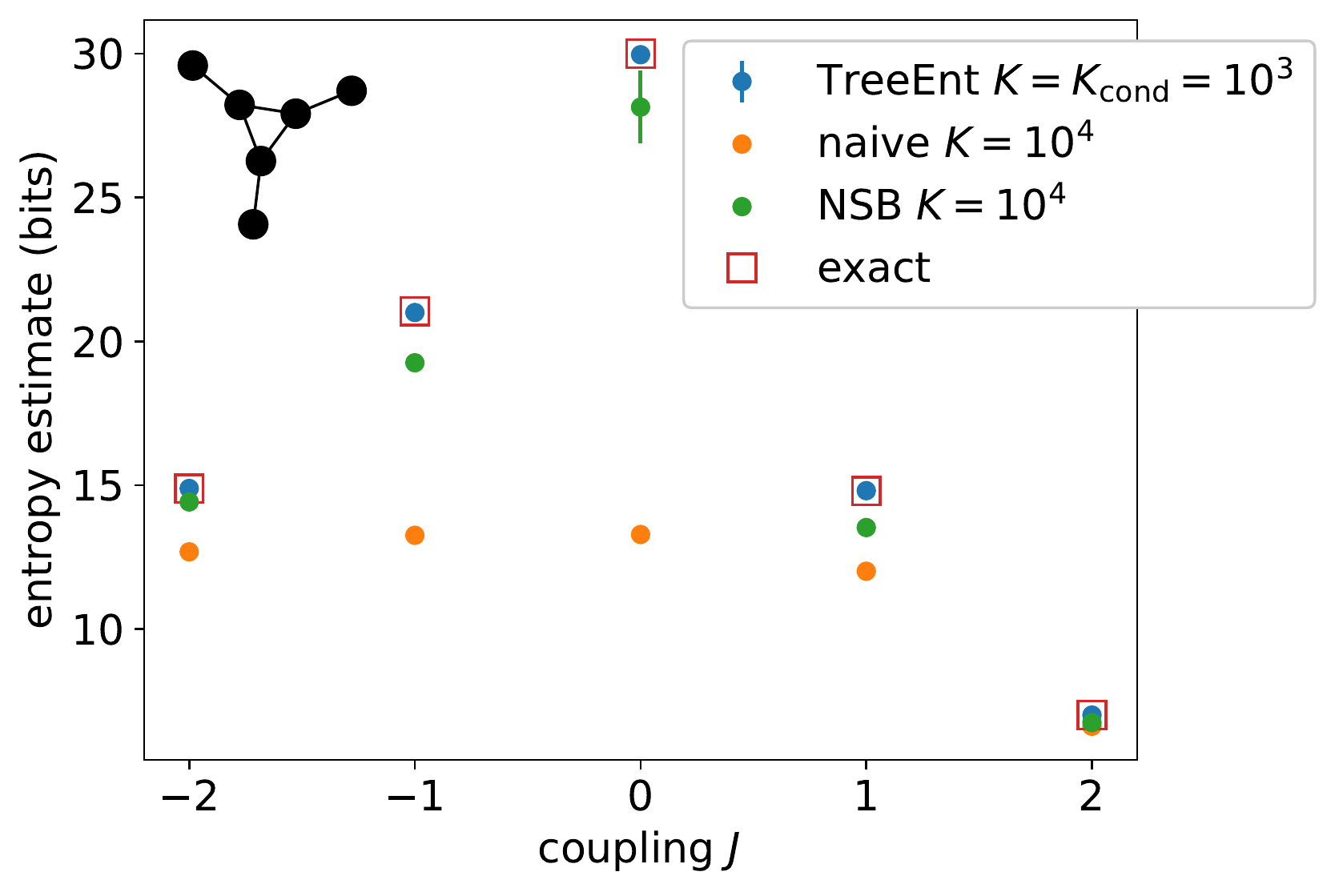}
	\caption{Comparison of TreeEnt with na\"ive and NSB entropy estimators on five replicas of an ``pointy'' triangle (inset on top left). Our algorithm does better faster and with fewer samples by recognizing the interaction structure of the model.}\label{gr:performance}
\end{figure}

\begin{figure*}[t]
	\includegraphics[width=.8\linewidth]{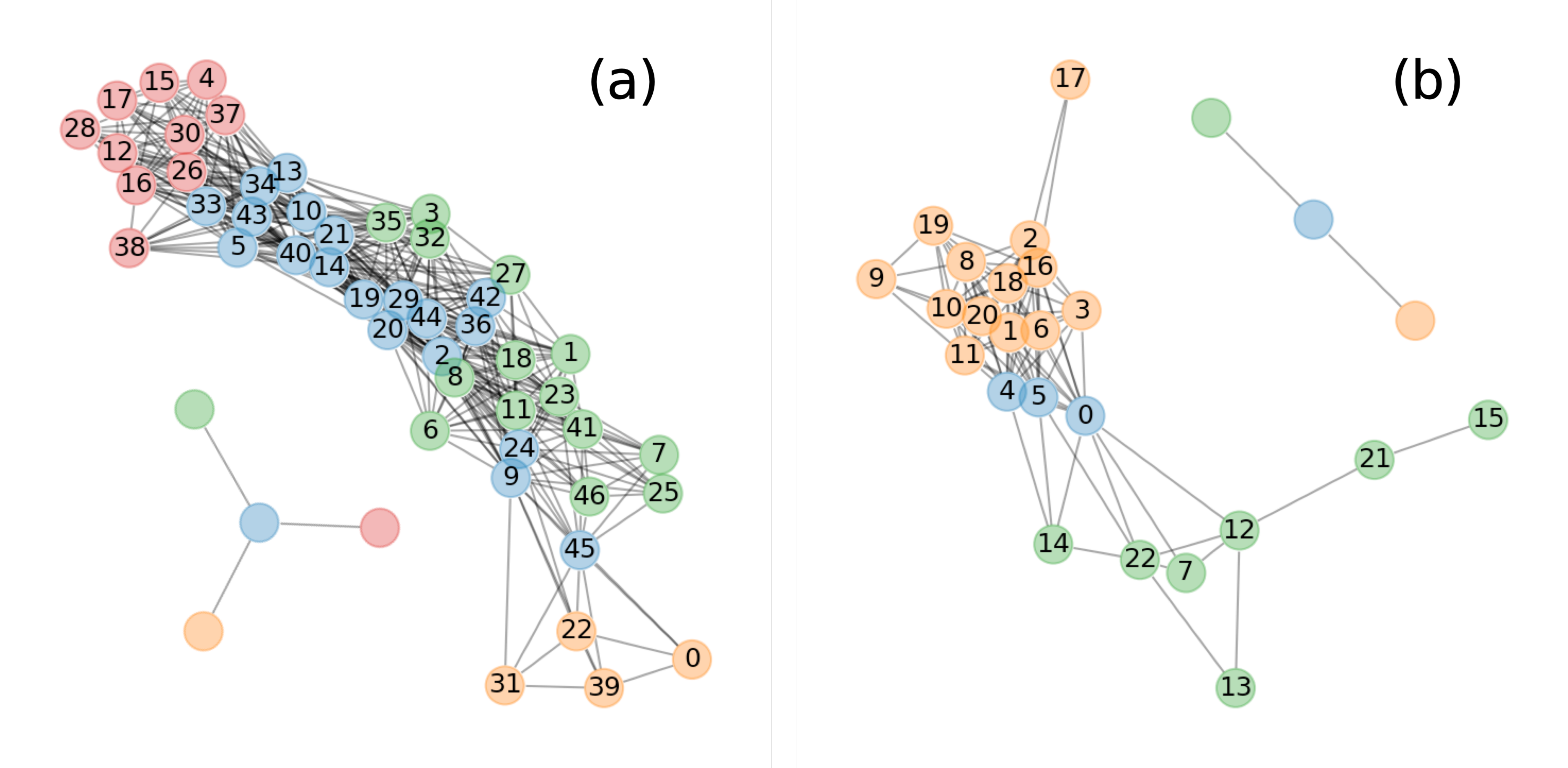}
	\caption{Examples of graph partitions using the minimal constraint heuristic applied to sparse US Appeals Courts interaction graphs from reference \citenum{}. (a) DC Circuit Court separates into four connected components when max cluster size is set to $M=14$. Largest component is of size $N=19$ compared to $N=47$ for the entire graph. Each color is a different subgraph. Inset shows interaction structure on coarse-grained graph which is a tree, i.e.~the green, red, and orange clusters are independent of one another once fixing the blue cluster. (b) First Circuit is partitioned into sets that form a line. Largest component is of size $N=13$ vs.~total graph size of $N=23$.}\label{gr:example}
\end{figure*}

\section{Error estimation}
For the algorithm, we must account for two sources of error including from the finite sample size of the conditioned set and the NSB estimator. 

For the finite sample contribution, the contribution to the entropy for a single term (the entropy of a leaf) in Eq~\ref{eq:recursive H} is the combination of terms 
\begin{align}
	\br{H[p(s_{i+1}^{a''})]} &= \sum_{s_{i}^{a'}} p(s_{i}^{a'}) H[p(s_{i+1}^{a''}|s_{i}^{a'})]\label{eq:mean}
\end{align}
where the sum is over all the unique states that occur for the branch $s_i^{a'}$ and for the particular leaf $a''$. For this calculation, we are not considering the sum over all the parents of the branch $s_i^{a'}$ because we can MCMC sample directly from the subset. This means that we can also calculate the variance in the entropy of this leaf, 
\begin{align}
	\sigma^2_{s_{i+1}^{a''}} &= \sum_{s_{i}^{a'}} p(s_{i}^{a'}) \Big(H[p(s_{i+1}^{a''}|s_{i}^{a'})] - \notag\\
		&\qquad\qquad\qquad\qquad H[p(s_{i+1}^{a''}|s_{i}^{a'})]\Big)^2.\label{eq:var}
\end{align}
As in the usual sense, we normalize the variance by the number of samples $K$ to estimate the standard error of the mean. Since the NSB error $\sigma_{\rm NSB}(s_i^{a'})^2$ on each term in the sums of Eqs~\ref{eq:mean} and \ref{eq:var} are computed independently of the finite-sample error, we add the norms of both the errors together to determine the total error on the entropy estimate. For a single leaf, we have the sum of the variance, or the norm-squared error
\begin{align}
	\Sigma_{s_{i+1}^{a''}}^2 &= \sigma^2_{s_{i+1}^{a''}}\Big/K + \sum_{s_i^{a'}} p(s_i^{a'})\sigma_{\rm NSB}(s_i^{a'})^2.
\end{align}

\section{Assessing performance}
We compare our algorithm with the na\"ive and NSB estimator in an example where the entropy can be exactly calculated by enumeration of all configurations. We find that leveraging the factorized structure of the graph allows us to recover a nearly perfect estimate to the exact entropy. As an example, we show the results for an ensemble consisting of multiple, disconnected graphs, each one consisting of a triangle with a single additional node coming off of each vertex (thus a ``pointy'' triangle) for a total of six spins. The spins are coupled with uniform strength $J$ that we vary. When we sample from five replicas of the graph at the same time, the state space of $K=2^{30} \sim 10^9$ is difficult to sample well on a desktop machine. Our estimator, as we show in Figure~\ref{gr:performance}, performs well across different scales of the coupling $J$ even with a relatively small number of samples. It is also substantially faster than applying the NSB estimator to the entire ensemble at once because the subsets on which we calculate the estimator are smaller.

\section{Application to maxent voting model}
We use our algorithm to estimate the entropy of a sparse voting model of judge voting on the US Circuit courts. The probabilistic models that we consider derive from the maximum entropy principle, which is an algorithm for determining minimal statistical models of data \cite{jaynesInformationTheory1957}. Importantly, the quantity of interest in this framework is the entropy of the model, which is maximized while constraining a few, crucial properties of the system such as the average vote of each judge and their pairwise correlations \cite{leeConvenientInterface2019}. When these two sets of constraints are imposed, we obtain the pairwise maxent model, which has been shown to model to high accuracy voting patterns on the US Supreme Court \cite{leePartisanIntuition2018, leeStatisticalMechanics2015}. 

According to the model, the probability distribution takes the Boltzmann form
\begin{align}
	p(s) &= \frac{1}{Z}e^{-E(s)}\label{eq:maxent p}
\end{align}
with normalization term $Z$ known as the ``partition function'' in statistical physics. To each configuration $s$ defined as a vector of $-1$ and $1$ for the votes of the judges, an ``energy'' $E(s)$ is assigned, where lower energy implies that the configuration is more likely. In the case of the pairwise maxent model, it takes the form
\begin{align}
	E(s) &= -\sum_{\rm i=1}^N h_{\rm i}\si - \sum_{\rm i<j}^N J_{\rm ij}\si\sj.\label{eq:E}
\end{align}
The fields $h_{\rm i}$ describe a bias for each component i such that a component with negative bias $h_{\rm i}<0$ will tend to take on a value of $-1$, whereas one with positive bias $h_{\rm i}>0$ will tend to take on value of 1. The couplings $J_{\rm ij}$ describe the tendencies of components to align with each other. Similar to the bias, a positive coupling lowers the energy when two components are aligned and a negative coupling lowers the energy when the components are misaligned. Thus, the pairwise maxent model captures both an independent tendency to be biased in one direction or another and a pairwise interaction tendency for how one component interacts with every other.

\begin{figure}
	\includegraphics[height=.65\linewidth]{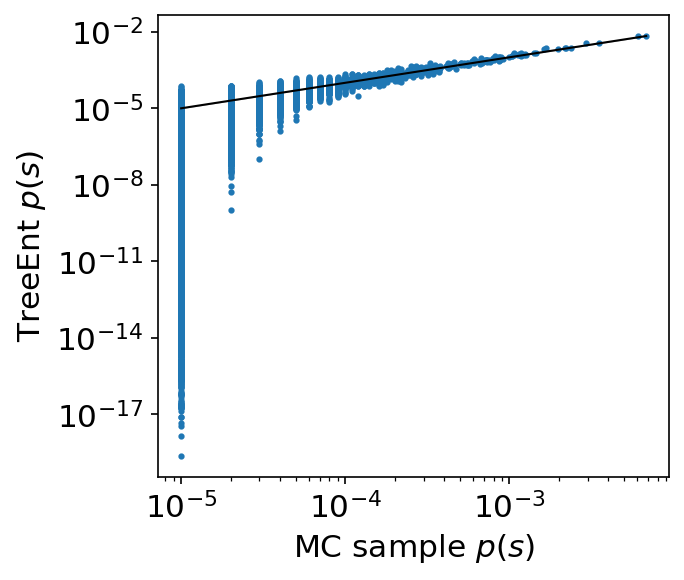}
	\caption{Probability of state $s$ using our entropy estimate for the partition function vs.~from Monte Carlo Markov chain sampling for the DC Circuit. We show points that appeared at least twice in the sample. As in Eq~\ref{eq:free energy}, we must also estimate the average energy $\br{E}$, which is straightforward to obtain with high precision using an MCMC sample unless the distribution of the energy displays a heavy tail such as near a critical point---but this will also pose a problem for the entropy estimate. MCMC sample size of $K=10^6$.}\label{gr:dc p}
\end{figure}

Importantly, the couplings describe a statistical interaction network akin to the edges displayed in Figure~\ref{gr:graph}. This is visible from looking at the factorization of the probability distribution described in Eqs~\ref{eq:maxent p} and \ref{eq:E}. It is clear from the additivity of the energy function that a separate term in the product for $p(s)$ appears for every pair of voters i and j related by $J_{\rm ij}\neq 0$. Thus, we only need use the connectivity of the matrix of couplings to break the graph into conditionally independent components.

In the US Court of Appeals, the corresponding pairwise maxent model is sparse because the number of judges sitting on a court is capped at any given time. This means that many judges in the past have had no interaction with judges in the future and therefore there is no coupling between them (more details see the supplementary information in reference \citenum{leePartisanIntuition2018}). As we show in Figure~\ref{gr:example}a, our algorithm factorizes the interaction graph for the District of Columbia (DC) circuit into four components: a simple tree, where the green, red, and orange sets are rendered conditionally independent given the blue cluster. While the na\"ive approach would have required estimating in a state space corresponding to $N=47$ voters, or of size $2^{47}\sim 10^{15}$, the largest cluster after our partition is of $N=19$, or a state space of about $10^6$, which is feasible to sample well numerically and for which we expect the NSB estimator to work well with many fewer samples than the full state space. In Figure~\ref{gr:example}b, we show the interaction graph of the First Circuit, where $N=23$, and we again find a partition into three clusters, two of which in green and orange are conditionally independent of one another once fixing the center blue cluster. Thus, TreeEnt provides an accelerated and more accurate method for estimating the entropies of these voting models.

\begin{figure}
	\includegraphics[height=.65\linewidth]{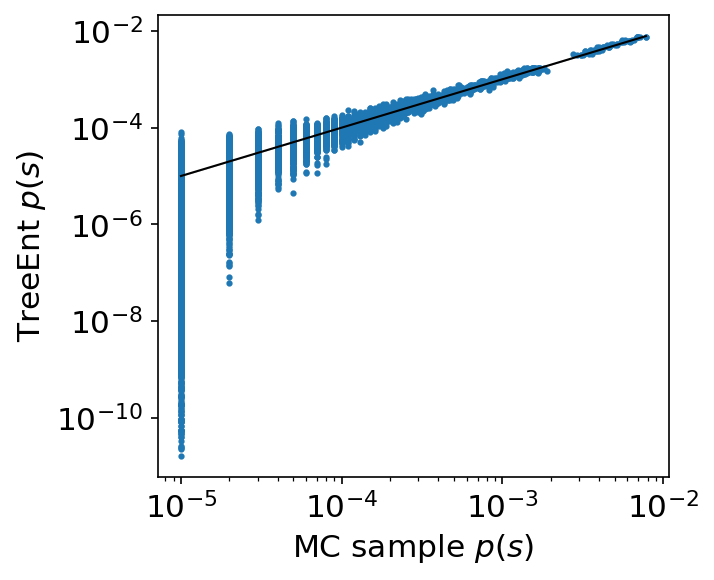}
	\caption{Probability $p(s)$ of state $s$ using our entropy estimate for the partition function vs.~from MCMC sampling for the First Circuit. See Figure~\ref{gr:dc p} for more details. MCMC sample size of $K=10^5$.}\label{gr:first p}
\end{figure}

For the best fit models, our estimator finds an entropy of $S_{\rm DC}=17.84\pm 0.05$\,bits for the DC Circuit and $S_1=11.34\pm0.03$\,bits for the First Circuit once having set branch sample size $K=10^4$ and leaf sample size $K'=10^3$. Since the entropy estimates represent substantial decreases in the entropy from the independent model of $S_{\rm DC}=47$\,bits and $S_1=23$\,bits, they show that the voting behavior of the judges is consistent with strong correlations. In comparison, the pairwise maxent model of the US Supreme Court from 1994--2005 is about 5 bits out of the possible maximum of 9 bits. In terms of the entropy per voter, the appeals courts are more constrained than the US Supreme Court, which is sensible because they are generally obligated to follow precedents set by the latter.

The entropy is intimately related to the partition function, allowing us to estimate the partition function from the entropy in a reversal of the usual procedure. Starting with the logarithm of Eq~\ref{eq:maxent p},
\begin{align}
\begin{aligned}
	\log p(s) &= -E(s) - \log Z, \\
	-\log Z &= \br{E} - H[p],\label{eq:free energy}
\end{aligned}
\end{align}
where in the last step we took the average of both sides over the distribution $p(s)$. This leads to the Helmholtz free energy relation in Eq~\ref{eq:free energy}. Thus, we can use our estimate for the entropy to calculate the partition function, which is an important quantity that, besides determining the normalization, is directly involved in calculations of physical properties of the model. As we show in Figures~\ref{gr:dc p} and \ref{gr:first p}, we obtain excellent agreement with the probabilities of states $p(s)$ estimated from MCMC sampling using this estimate of the partition function to normalize the probabilities.

\begin{figure}\centering
	\includegraphics[width=\linewidth]{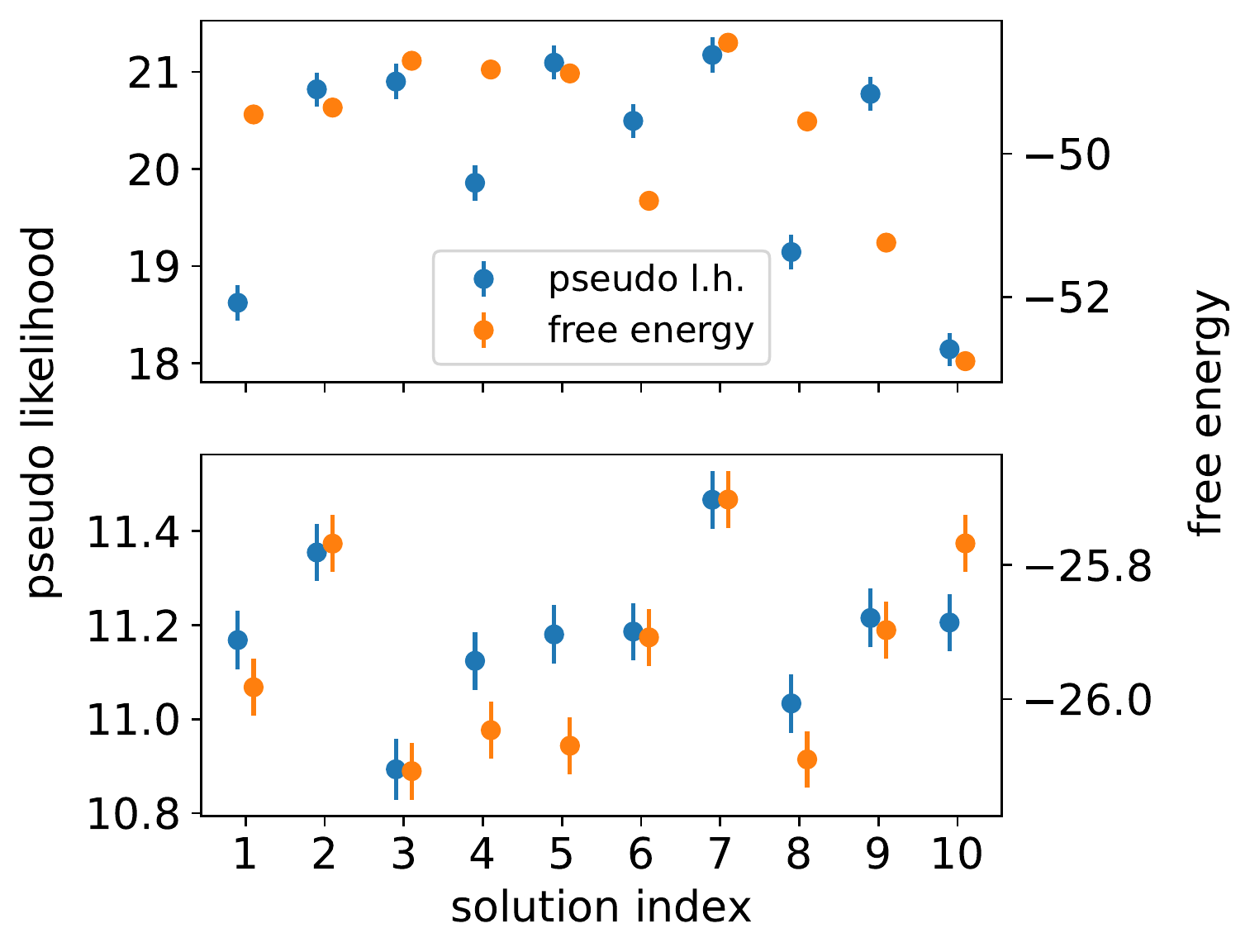}
	\caption{Goodness of fit measured by pseudo likelihood (blue and defined in Eq~\ref{eq:pseudo dkl}) and free energy (orange) for ten different, degenerate solutions to the DC and First Circuits. The collective measure reveals some of the solutions to be superior to others along both counts. Errors from entropy estimation as described in the main text and standard errors of the mean from energy estimates are summed by norm (the squared errors are summed and then a square root taken).}\label{gr:dkl}
\end{figure}

In the case of the solution landscape for the Court of Appeals, the solution landscape is degenerate because of the set of consistency conditions used to impute missing votes \cite{leePartisanIntuition2018}. As a result, we require a way of choosing amongst the multiple solutions that we recover. Each solution has the same interaction structure, which is determined by the pairs of judges we have observed voting together or not voting together, but the particular values of the fields $\hi$ or couplings $\Jij$ will change. As a standard information theoretic measure of the goodness-of-fit, we rely on the KL divergence between the distribution of the data and the model,
\begin{align}
	D_{\rm KL} &= \sum_{s} p_{\rm data}(s)\log\left(\frac{p_{\rm data}(s)}{p(s)}\right).
	\intertext{This reduces to a ``pseudo likelihood'' after ignoring the entropy of the data, which is a constant that does not matter for minimizing the divergence, or}
	\tilde D_{\rm KL} &\sim \br{E}_{p_{\rm data}} + \log Z.\label{eq:pseudo dkl}
\end{align}
We call Eq~\ref{eq:pseudo dkl} a pseudo likelihood because it leads to the same relative outcome as when maximizing the likelihood. Note that we have inserted the form for the maxent model for $p(s)$. Then, the first term is the average energy weighted by the data distribution (once marginalized over any unobserved voters) and the second the free energy as given in Eq~\ref{eq:free energy}. As an alternative measure for goodness of fit, we can consider the free energy alone because it balances the energy (which more constrained distributions minimize) and the entropy (which more random distributions and thus of higher multiplicity maximize). As we show in Figure~\ref{gr:dkl}, these measure distinguish certain solutions, and the ones of best fit agree with qualitative checks on the ranked order of judges according to the imputed correlation matrix \footnote{We use the principal dimension of the imputed correlation matrix to define a relative ranking of judges, and the ranking is consistent with qualitative checks with a legal scholar on a conservative-liberal dimension.}.

Importantly, the likelihood is difficult to compute, and as a result a typical way to assess the fit of maxent models is to use correlations that have not been explicitly fed into the model. For example, the Ising model should reproduce exactly the mean vote and the pairwise correlations, but there is no guarantee that it can reproduce higher-order correlations. These statistics are often used for quality of fit in models of neural statistics and voting \cite{leeStatisticalMechanics2015, schneidmanWeakPairwise2006}. In voting data sets, relying on such checks poses serious problem when votes are missing such as when only subsets of individuals vote together (e.g.~a standard bench is only of three judges in the appeals courts) and checking correlations is not possible. Here, we show that it is feasible to do an accurate and better comparison between models using our heuristic.

\section*{Discussion}
We propose a heuristic algorithm for calculating an essential collective property of graph models of social behavior, the information entropy. In contrast with the usual estimates of collective properties and measures of goodness-of-fit with prediction of lower-order statistics \cite{leeStatisticalMechanics2015,schneidmanWeakPairwise2006}, we show that leveraging sparse graph structure along with good estimators of the entropy can allow us compute collective statistics that implicitly incorporate correlations of all orders. As examples of our algorithm, we present two judicial voting models for the US Court of Appeals. We find that we can obtain precise estimates of the entropy that we then use to identify an optimal solution from an ensemble of degenerate solutions. From our calculation, we also discover that the relative entropy per judge is higher than that of the US Supreme Court, a useful observation step for further work on comparison of institutional properties and constraints from behavioral data.


For the problems that we consider, the sociological measure for detecting bridges, the structural contraint work well, but our work could be extended by considering alternative algorithms for factorizing the graph. In experimenting, we found that other common measures such as between centrality or Louvain community clustering were not as good at factorizing the graphs for our recursive entropy calculation. One potential fruitful direction would be to consider structural holes of higher order. Furthermore, incorporating effective resistance as edge weights for the structural constraint did not improve matters, although surely the success of an algorithm will depend on the properties of the graph. As a step towards future work, our current work provides a useful step in assessing both collective properties and model fit in the context of graph models for social interaction.

\section{Architecture}
An open-source package for our algorithm TreeEnt will be available on \url{https://github.com/eltrompetero/treeEnt}, where TreeEnt is shorthand for Tree Entropy (as well as a reference to Lord of the Rings, where mobile trees are known as Ents).

The core of the package consists of two Python modules contained in ``measures.py,'' ``test\_measures.py,'' and ``NSB\_toolbox.py''. As the names indicate, the algorithm described in the main text is implemented in the measures module as part of the TreeEntropy class. The testing module is contained in the test module that provides some automated tests that can be run with pytest as well as routines for checking the validity of the algorithm as is implemented in the accompanying Jupyter notebook. The final module contains an implementation of the NSB estimator, which is heavily borrowed from an existing codebase written by Bryan Daniels at \url{https://github.com/bcdaniels/toolbox.git}. 

The TreeEnt class takes an model instance that describes the probabilistic graph model. This instance must have routines for sampling from the probabilistic model of interest. In the current implementation, we base this class on the Ising model class implemented in the ConIII coding project. Running the computation with ConIII entails a much larger number of dependencies than those explicitly identified as part of TreeEnt.

\section{Acknowledgements}
EDL acknowledges funding from the Austrian Science Fund under grant number ESP 127-N. We acknowledge useful discussions with Cris Moore and Bryan Daniel's help with his open-source code.

We declare no competing interests.

\bibliography{refs}

\end{document}